# Electron residual energy due to stochastic heating in field-ionized plasma

Elnaz Khalilzadeh[1,2], Jam Yazdanpanah[2], Jafar Jahanpanah[1], Amir Chakhmachi[2], and Elnaz Yazdani[3]

[1]Department of Physics, Kharazmi University, 49 Mofateh Ave, Tehran, Iran

[2]The Plasma Physics and Fusion Research School, Tehran, Iran

[3]Laser and optics research school, Tehran, Iran

**Abstract**

The electron residual energy originated from the stochastic heating in under-dense field-ionized plasma is here investigated. The optical response of plasma is initially modeled by using the concept of two counter-propagating electromagnetic waves. The solution of motion equation of a single electron indicates that by including the ionization, the electron with higher residual energy compared to the case without ionization could be obtained. In agreement with chaotic nature of the motion, it is found that the electron residual energy will significantly be changed by applying a minor change to the initial conditions. Extensive kinetic 1D-3V particle-in-cell (PIC) simulations have been performed in order to resolve full plasma reactions. In this way, two different regimes of plasma behavior are observed by varying the pulse length. The results indicate that the amplitude of scattered fields in sufficient long pulse length is high enough to act as a second counter-propagating wave for triggering the stochastic electron motion. On the other hand, the analyses of intensity spectrum reveal this fact that the dominant scattering mechanism tends to Thomson rather Raman scattering by increasing the pulse length. A covariant formalism is used to describe the plasma heating so that it enables us to measure electron temperature inside the pulse region.

I - Introduction

The rapid advances in the generation and amplification of ultra-short laser pulses [1, 2] has led to the laser–plasma interaction to be one of the most remarkable topics in physics. In this context, understanding of heating mechanisms in an under-dense plasma has received interest due to their potential applications, such as fast ignition in ICF [3, 4], proton acceleration [5, 6], x ray lasers [7, 8], and laser wakefield accelerators [9, 10]. When an intense laser pulse passes through a gas, the plasma can be formed by field ionization. The plasma is normally formed by the field ionization during passing an intense laser pulse through a gas system. In this process, the electrons are driven toward the both fields of laser and space charges induced by the electron motion. The net energy that transfers to the electrons in the presence of laser field by the different heating mechanisms is known as electron residual energy. Recently, the electron residual energy in optical field-ionized plasmas has extensively been elaborated by many theoretical and experimental physicists [11-18]. The residual momentum and energy of electron are analytically investigated as a function of gas and laser parameters. It has been found that the residual energy can be reduced by space-charge effect [19].The harmonic oscillator model has then been implemented to describe the electron motion in a laser field, in which the space-charge field acts as a restoring force. However, He et al. [20] were shown that in moderate density conditions, the space-charge field can increase the residual energy of electrons effectively by matching the space-charge field with laser pulse. When the electron density and the laser pulse length were increased, it is found that the inverse bremsstrahlung heating and stimulated Raman scattering play the important role in plasma heating [21, 22].

Other important mechanism which led to the effective heating of plasma electrons is stochastic heating in the presence of two counter-propagating laser pulses. It is obvious that the single electron motion in a plane electromagnetic wave is integrable [23]. However, in presence of another plane wave, the corresponding Hamiltonian is usually not integrabl. In this case, the electron dynamic becomes chaotic when the second pulse intensity passes a certain threshold. Therefore, it is not possible to evaluate the maximum kinetic energy of electrons because they are heated to an energy level much larger than that of a single plane wave. The thresholds of local stochastic motion have been estimated by Mendonca [24] as $a_1 a_2 = 1/16$ (that $a_1$ and $a_2$ are the vector potential). Many theoretical and experimental evidence of stochastic heating have been displayed using two crossed laser pulses. Using WAVE code simulations, Leeman et al. have shown that the stochastic heating play an important role in plasma heating [25]. The particle-in-cell-simulation reveals that the stochastic heating is a dominant mechanism in semi-infinite laser interaction for the moderate plasma densities, such as a few percent of the critical density [26, 27]. It is demonstrated that the counter-propagating waves can either be the Raman backscattered wave or the reflected wave from high dense plasma.

In this work, the electron residual energy due to stochastic heating in the presence of the counter-propagating electromagnetic waves (limited main pulse + backscattered) by means of single electron calculations and particle-in-cell simulation is studied. The field-ionization process is modeled for the Hydrogen atoms interacting with main laser pulse at the initial step. In addition, the ionization algorithm is improved by considering the Stark effect and non-zero velocity for ionized electrons. The result of single electron calculations is shown that the residual energy strongly dependents on the initial conditions of electron when it interacts with the second pulse. Therefore, the laser energy transformation to the electron is significantly changed by applying any small change to the initial conditions. This behavior is due to the electron stochastic motion.

A comprehensive research of different electron heating regimes is here managed by employing the fully kinetic 1D-3V particle-in-cell simulation, and by choosing two finite laser pulse lengths with the laser intensity $I = 10^{18} w/m^2$ and the density of Hydrogen atoms $n = 0.01\ n_{cr}$ ($n_{cr}$ is critical density). It is noticed that the field-ionized plasma strongly reacts against the laser field oscillations in order that the scattered fields are produced. Furthermore, the amplitude of scattered fields is increased by extending the pulse length which in turn led to the different electron heating regimes. Two finite laser pulse lengths $\tau_L = 100\ fs$ and $\tau_L = 250\ fs$ are typically chosen to describe the physics represent by this different regime. It is illustrated that unlike short pulse ($\tau_L = 100\ fs$), in the case of long pulse ($\tau_L = 250\ fs$), the amplitude of scattered fields is high enough to act as a second counter-propagating wave for initiating the stochastic electron motion. In the most previous works [28-30], the scattered fields have been attributed to Raman scattering whereas our simulation results show that the type of scattering mechanism is depending on the laser pulse duration in the same initial conditions. However, the dominant mechanism of Raman scattering is substituted by Thomson scattering by extending pulse length. In our work, a covariant formalism is used to describe the plasma heating and to measure electron temperature inside the pulse region.

The paper is organized according the following sections. Sec. II contains a general presentation of theory. In Sec. III, we describe our numerical PIC code. The conclusion is given in Sec. IV.

## II- The single electron dynamics

We start by considering the Hydrogen atom in the main laser pulse. When an atom is exposed to an ultrahigh optical intensity, electrons can be liberated from the atom through different mechanism. There are a considerable number of studies have been devoted to the investigation of ionization. The basic concepts were developed by Keldysh [31], who introduced the parameter $\gamma = \omega_L \sqrt{2I_p}/E$, where $I_p$, $\omega_L$, and $E$ are the atomic ionization potential, and the frequency and amplitude of the laser field, respectively (atomic units is used throughout this section). Keldysh shown that $\gamma \ll 1$ correspond to the Tunneling ionization (TI). In this case, when the laser electric field is strong enough, such that $E > E_{critical}$, (Bauer [32] has shown that this critical value for atomic Hydrogen is given by $E_{critical} = (\sqrt{2}-1)|I_p|^{3/2} = 0.146$ a.u) the potential barrier is suppressed below the electron energy level, freeing the electron. This ionization mechanism is so called Barrier suppression ionization (BSI). Bauer and Mulsner [33] have proposed a combined formula (a combination of theoretical and empirical formula) which is used in this paper for the region $E < 0.175$ as

$$w = \begin{cases} \dfrac{4}{E}\exp(-\dfrac{2}{3E}) & \text{for } E < 0.084 \quad \text{(Landau formula)} \\ 2.4\ E^2 & \text{for } 0.084 \leq E < 0.175 \quad \text{(emprical formula)} \end{cases} \qquad (1)$$

By applying very strong electric fields, numerical calculations in a constant electric field shows that the values of the Stark shift associated with the ground state of Hydrogen atom are strongly diminished by increasing the field strength which is in contrast with second-order perturbation theory. In this region, according to the calculations of Ref. [34, 35], the determined ionization rates can be nicely fitted by the following equation

$$w = 1.29\ E - 0.15 \qquad \text{for}\ \ 0.175 \leq E \leq 0.8\ . \tag{2}$$

However, noting to Eqs. (1) and (2), depends on the amount of the electric field strength in moment of ionization, the electron is ionized by the main pulse which propagates in the positive $x$ direction. Later on, the ionized electron moves in the presence of second pulse which propagates in the negative $x$ direction by certain delay space with respect to the main pulse. This delay space is determined by the parameter $D_x$. In many studies, it has been shown that electron dynamics in a strong laser pulse can change dramatically in the presence of another relatively weak, counter-propagating second laser pulse. When the second pulse intensity passes a certain threshold the electron dynamics becomes chaotic. To accurate study of this mechanism, a numerical solution of single electron motion equations, for two contra-propagating pulses, using of Monte Carlo method is examined. It is assumed that the electric fields of the laser pulse have a Gaussian shape.

It is evident that the momentum of the electron in a main pulse is determined by $p_y + a_1 = C_1$ and $p_x - \gamma = C_2$ [36], where $p_x$, $p_y$, $a_1$, and $\gamma$ are the longitudinal and transverse momentum (normalized by $mc$), vector potential (normalized by $mc^2/e$) and relativistic gamma factor, respectively. And $C_1$ and $C_2$ are constant. The solution with zero initial velocity of the ionized electron is given by

$$p_x = \frac{(a_0 - a_1)^2}{2}, \quad p_y = -a_1 + a_0, \tag{4}$$

where $a_0$ is the initial vector potential of the main pulse in the moment of ionization. Figure 1 shows the trajectories of a test electron in momentum phase in vacuum (a) with a single plane electromagnetic wave, $a_1 = 2$, and (b) two counter-propagating electromagnetic waves, $a_1 = 2$ and $a_2 = 0.2$. The parabolic curve between $p_x$ and $p_y$ is obtained in Fig. 1(a). In presence of second counter-propagating wave, the electron trajectory in the momentum space is no longer confined to the parabolic curve. Instead, it spreads in the momentum and creates the region for stochastic motion in momentum space, as shown in Fig. 1(b).

It is clear that the ionization of atom can occur at the different locations and times of main laser pulse. According to the Eq. (4), the amount of main pulse electric field (vector potential) at moment of ionization, $a_0$, and at moment of interaction with second pulse, $a_1$, (support to $D_t$ variation) specify the initial conditions for the ionized electron when it interacts with the second pulse. Since the stochastic motion of electron in the presence of the counter-propagating electromagnetic waves is strongly dependent to the initial conditions, it is expected that by initial condition changing, the electrons with different residual energy are attained. This subject is studied in the figure 2 and 3.

In figure 2, time history of electron energy for peak amplitude $a_1 = 2$, the second counter-propagating pulse at a peak amplitude $a_2 = 0.3$ and $D_x = 10\mu m$ is shown. The result of Fig. 2(a) without including the atom ionization is obtained, where the residual energy of electron is 23 keV. In figures 2(b)– 2(d), three electrons with three different $a_0$ (these electrons are ionized in three different locations and times of first pulse) are considered. Electrons with different residual energy are obtained as expectedly. In addition, it is important to notice that the electron residual energy by including the ionization can be higher than the case without ionization.

The effect of $D_x$ variation on the electron residual energy is shown in Figure 3, for $a_1 = 2$, and $a_2 = 0.3$, with the same $a_0$ (these electrons are ionized in the same location and time of main pulse) for different $D_x$ as (a) $8\mu m$, (b) $9\mu m$, (c) $10\mu m$, and (d) $10.3\mu m$. Based on the results, by small changes in the initial conditions significant changes in the electron residual energy is observed. This can only be attributed to the stochastic motion which is strongly dependent to the initial condition.

However, in this section the physical description of stochastic motion of the single electron is presented based on the simple model without considering the space charge effect. In the next section, we use particle simulation to obtain precise results for the electron heating with considering the collective effects of plasma formed due to the ionization of atoms.

III- simulation results and discussion

To avoid the repeating of simulation parameters through this section, these parameters are first presented before the simulation results. Using PIC code which was developed by J. Yazdanpanah [37-38] (see also [39-40]), we have carried out extensive 1D-3V PIC simulations for laser wavelength $\lambda = 1 \mu m$ and laser intensities $10^{18}$ Wcm$^{-2}$ ($a_0 = 1$). For all run-instances, we have used hydrogen atoms with density $n_H = 0.01$ $n_{cr}$ ($n_{cr}$ is critical density) and with initial profile that have been invariantly step like. The shape of laser pulse has been assumed to be such that the pulse envelope rises (with sinus functionality), then remains constant and finally falls symmetrically. High spatial resolution of 200 cells per laser wavelength with at least 64 particles per cell is used in the simulations. The spatial resolution guaranties the plasma against the un-physical heating produced by the finite-grid instability by appropriate resolution of the Deby length $\lambda_D$, i.e.

$DX/\lambda_D \approx 0.3$ [38-39]. Reflecting and open boundary conditions have been applied for particles and fields respectively.

In order to study the electron heating mechanism, the longitudinal momentum, the transverse vector potential and the longitudinal electric field versus x, for $\tau_L = 100 fs$ at times (a) $175 fs$, (b) $437 fs$, and (c) $700 fs$ are plotted in Fig. 4. The atoms start to ionize as soon as the laser pulse interacts with the atoms and the plasma is forming. This plasma reacts against the pulse field oscillations and the scattered field is produced. With attention to detail in Figure 4, it is found that the amplitude of the backscattered wave is low to drive the electron stochastic motion. And the nonlinear wave breaking [41] is only mechanism which leads to the electron heating. It is expected that with increasing of length pulse, the amplitude of scattered fields increase and different electron heating regime is happened. To describe the physics represented by these regime, the simulation result for $\tau_L = 250 fs$ is given in Fig 5. In this figure, the transverse vector potential and the longitudinal electric field versus x are shown at times (a) $175 fs$, (b) $437 fs$, and (c) $700 fs$. The scattered radiation and considerable changes in the electron phase space are evident. With attention to detail in Figure 5, it is found that two important mechanisms might be described the electron heating inside the disturbed plasma, which in turn produce two different electron configurations in phase space. The first configuration consists of electrons produced due to self-modulation wave breaking. Self-Modulation between radiation waves and scattered waves produces a pondermotive force that it can derive the plasma longitudinal electrostatic mode. As a result, the amplitude of the plasma oscillations increases efficiently and the necessary conditions for wave breaking is provided [42-46]. By the time interaction progressing, the second electron pattern is appeared in phase space. Since the self-modulation of scattered and incident waves and superposition between the scattered waves are taken place, the amplitude of scattered fields is high enough to act as a second counter-propagating wave and trigger the stochastic electron motion. Therefore, it leads to transfer effective energy from laser fields to the electrons. As expected, in the latter case, the electrons travel between different attractor while the electron trajectory in phase space becomes irregular. However, in the first case, the electrons travel among different attractor whereas their trajectories remain regular. It should be noticed that the main mechanism for electron heating is stochastic heating. As a result, it is obvious that the pulse length has an essential role in stochastic heating and should consider as an important parameter in this type of heating studies.

More explanation of the physical insight of scattering mechanisms in above outcomes is as follows: the scattering of electromagnetic radiation by a free charged particle is so called Thomson scattering. On the other side, when the plasma collective effects are important, the mechanism of scattering goes to Raman scattering. Since Thomson scattering cross section is several times larger than Raman scattering cross section, the ratio of the bunched electron density to the plasma density plays an effective role in dominant scattering occurrence. In the case of $\tau_L = 100fs$ in Fig. 5 ($\tau_L = 100fs$), it seems that the formation of dense regular plasma mode is the main reason in which Raman scattering becomes dominant mechanism. However, as seen in Fig. 4 for $\tau_L = 250fs$, different areas with dens electron bunches are appeared due to the wave breaking and stochastic motion of electrons. This could lead to Thomson scattering becomes as dominant scattering mechanism. In the following, to detail study of field scattering mechanisms in the Figs. 4 and 5, the intensity spectrum of total radiation (main field + scattered field) is plotted in Figs. 6(a) and 6(b) for pulse lengths $\tau_L = 100fs$ and $\tau_L = 250fs$, respectively. In the case of Fig. 6(a), the field modes with a definite shift in wavelength can be found by comparing of the field intensity spectrums at $t = 350fs$ and $t = 700fs$ with the initial incident pulse and there is also a specific phase relation between them. Therefore, this scattering mechanism could be attributed to Raman scattering. The total intensity spectrum in Fig. 6(b) indicates that there is no shift in the central wavelength of fields at $t = 400fs$ and $t = 800fs$ in comparison to the initial incident pulse. And, any particular phase relation between the total fields and the initial main fields is seen. Therefore, it seems that this scattering refer to Thomson scattering.

**Electron temperature measurement**

In the following, a new measurement method for plasma temperature is studied [38]. For this purpose, an infinitesimal fluid element is defined. Since plasma particles are affected by self-consistent electromagnetic forces, their energy is not a function of the electron temperature merely. In the relativistic-fluid formulation, the internal energy density ($e$) and fluid density ($n$) can be defined for fluid element, invariantly, as follow:

$$n = \int d^3p f(t,\mathbf{x},\mathbf{p})\Big|_{LR}$$
$$e = \int d^3p [f(t,\mathbf{x},\mathbf{p})mc^2\gamma]\Big|_{LR} \tag{1}$$

where $f(t,\mathbf{x},\mathbf{p})$, $\mathbf{p}$, $\gamma$, $m$, and $c$ are the electron distribution function, momentum, gamma factor, mass and speed of light, respectively. The LR index is refer to *local rest frame*, wherein the fluid is at rest at each time-space point $(t,\mathbf{x})$. With using this formulation, plasma temperature in different directions as an independent quantity from the plasma macroscopic motion which is affected by applied force, can be defined. The temperature can be written as $T_{ij} = P_{ij}/n_s$ where,

$$\mathbf{P}_{i,j} = \mathbf{P}_{j,i} = \int d^3p [f(t,\mathbf{x},\mathbf{p})p_i v_j]\Big|_{LR}$$

is the invariant components of the stress tensor and $n_s$ is the special density. The pressure tensor components in the quasi-static regime can be solved analytically. Here, we list the solutions obtained for an initially isotropic plasma temperature

$\theta \equiv k_B T_e / m_e c^2$, in the laboratory frame and in two velocity-space dimensions [38, 47],

$$\Theta^{1,3} = \Theta^{2,3} = \Theta^{3,3} = 0,$$

$$\frac{\Theta^{1,2}}{n_0 m_e c^2} = \left(\frac{n}{n_0}\right)^2 \frac{a_y}{\beta_g} \theta, \quad \frac{\Theta^{2,2}}{n_0 m_e c^2} = \frac{n}{n_0} \theta,$$

$$\frac{\Theta^{1,1}}{n_0 m_e c^2} = \gamma_u^2 \left(\frac{n}{n_0}\right)^3 \left[1 + \left(\frac{a_y}{\beta_g \gamma_u}\right)^2\right] \theta$$

that $\beta_g$, $n_0$, and $\gamma_u$ being respectively the pulse group velocity and unperturbed plasma electron density. The momentum-spread tensor calculated in the *local rest frame* is obtained from the momentum-spread tensor calculated in the laboratory frame via the double Lorentz transformation

$$P_{i,j} = \Lambda^i_\mu \Lambda^j_\nu \Theta^{\mu\nu}$$

where general form of the Lorentz transformation $\Lambda^\alpha_\mu$ in terms of the fluid velocity $\beta$ is given by

$$\Lambda^0_0 = \gamma$$
$$\Lambda^0_i = -\gamma \beta_i$$
$$\Lambda^j_i = \delta_{ji} + \frac{\beta_j \beta_i (\gamma - 1)}{\beta^2}$$
$$\Lambda^j_0 = -\gamma \beta_j$$

That $\gamma = (1 - \beta.\beta)^{-1/2}$ and $\Lambda^i_\mu$ are the fluid gamma factor and the Kronecker delta, respectively.

Here, the elements of stress tensor are independently calculated, without any particular assumption, using the PIC code. It should be noted that despite of plasma macroscopic motion, the introduced fluid element uses to measure the temperature inside and behind the laser pulse. Accordingly, changing of the electron heating amount in different pulse durations are compared and full information of the electron residual energy is obtained.

In Fig. 7, the simulation result for the stress tensor components, (a) $P_{11}$ and (b) $P_{22}$, for $\tau_L = 250 fs$ at $t = 700 fs$ are presented. These components provide a proper scale for electron energy in the different plasma regions. As shown in figure 7(a), when the laser pulse leaves the electrons, the relatively high energy electrons are appeared. Also it is obvious that, $P_{22}$ is small in the behind of laser pulse. This is because the Hamiltonian is independent of y and canonical momentum in y direction is constant of motion. Therefore, it is noted that a large portion of the electrons residual energy is due to the longitudinal momentum. Figure 8 displays the stress tensor components (a) $P_{11}$ and (b) $P_{22}$ at $t = 700 fs$ when the pulse duration is reduced to $\tau_L = 100 fs$. As expected, the low temperature electrons are generated in comparison with the electrons seen in Fig. 7.

IV- Conclusion

In this paper, the electron residual energy due to stochastic heating in the field-ionized plasma has been investigated in the presence of counter-propagating electromagnetic waves. When the main laser pulse interacts with Hydrogen atoms at initial step, the field-ionization process by considering the Stark effect and non-zero ionized electrons velocity has been investigated. The numerical solution of single electron motion equation has shown that by including the ionization, the electron with higher residual energy compared to the case without ionization could be attained. Also, by small variation in the initial conditions, the electron residual energy changes significantly.

To detail study of different electron heating regimes, 1D-3V particle-in-cell simulation for two finite laser pulse lengths $\tau_L = 100\ fs$ and $\tau_L = 250\ fs$ has been used. For $\tau_L = 100\ fs$, the amplitude of the backscattered wave is low to drive the electron stochastic motion. Therefore, the mechanism which leads to the electron heating is the nonlinear wave breaking. In the case of $\tau_L = 250\ fs$, it is found that two important mechanisms could be used to describe the electron heating inside of the disturbed field-ionized plasma. These mechanisms are self-modulation wave breaking and stochastic heating. In the latter case, the amplitude of scattered fields is high enough to act as a second counter-propagating wave and trigger the stochastic electron motion. Furthermore, based on the spectrum intensity analyses, it has been found that with increasing pulse length, dominant scattering mechanism approaches of Raman scattering to Thomson scattering. In this study, a new method which provides a proper scale for measurement of electron residual energy in different plasma regions has been used. Our results are also of interest for laser plasma scenarios involving production of high-energy electrons and protons due to heating and acceleration.

**FIGURE CAPTIONS:**

Fig. 1: Trajectory of a test electron in momentum phase in vacuum, (a) with a single plane electromagnetic wave, $a_1 = 2$, and (b) two counter-propagating electromagnetic waves, $a_1 = 2$ and $a_2 = 0.2$.

Figure 2: Time history of electron energy in the presence of two contra-propagating pulses with peak amplitudes $a_1 = 2$ and $a_2 = 0.3$ and $D_t = 10\mu m$, (a) without including the atom ionization, (b)–(d) three electrons with three different $a_0$.

Figure 3: Time history of electron energy in the presence of two contra-propagating pulses with peak amplitudes $a_1 = 2$ and $a_2 = 0.3$ and with same $a_0$, for different $D_t$ as (a) $8\mu m$, (b) $9\mu m$, (c) $10\mu m$, and (d) $10.3\mu m$.

Figure 4: The longitudinal electron phase space (left axis), the transverse vector potential (right axis), and the longitudinal electric field (left axis) versus x for $\tau_L = 100 fs$, at times (a) $175 fs$, (b) $437 fs$, and (c) $700 fs$.

Figure 5: The longitudinal electron phase space (left axis), the transverse vector potential (right axis), and the longitudinal electric field (left axis) versus x for $\tau_L = 250 fs$, at times (a) $175 fs$, (b) $437 fs$, and (c) $700 fs$.

Figure 6: the intensity spectrum of total radiation (main field + scattered field) is plotted for (a) $\tau_L = 100 fs$ and (b) $\tau_L = 250 fs$.

Figure 7: The simulation result for the stress tensor components, (a) $P_{11}$ and (b) $P_{22}$, for $\tau_L = 250 fs$ at $t = 175 fs$.

Figure 8: The simulation result for the stress tensor components, (a) $P_{11}$ and (b) $P_{22}$, for $\tau_L = 100 fs$ at $t = 175 fs$.

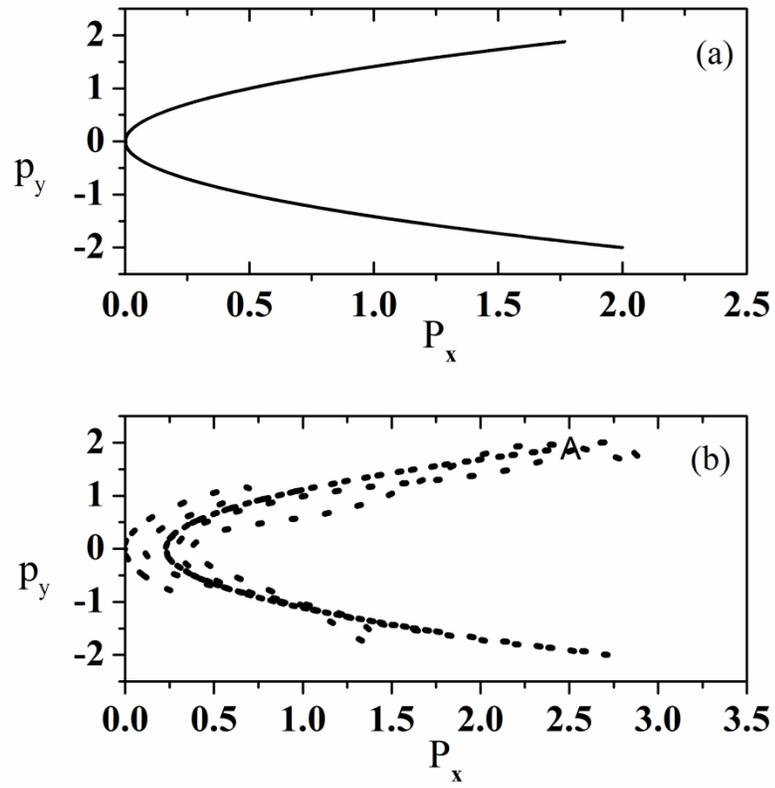

**Fig. 1**

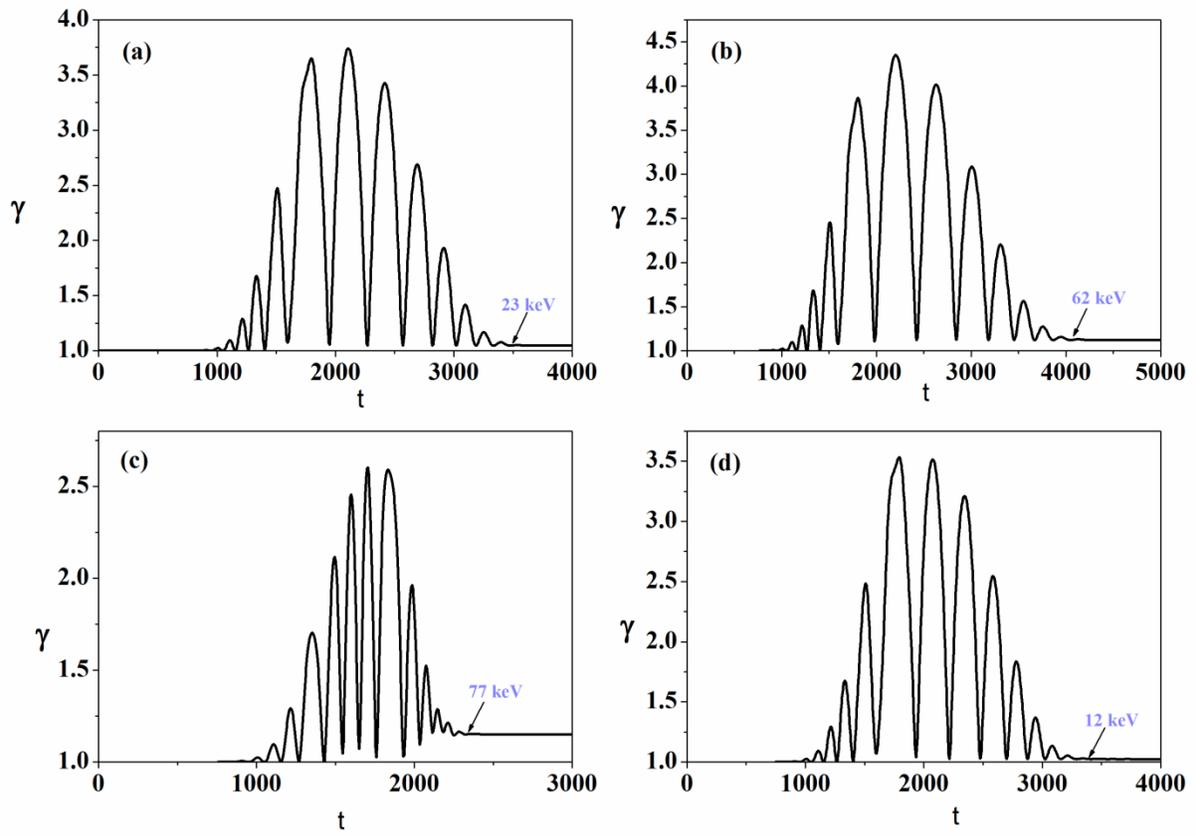

**Fig. 2**

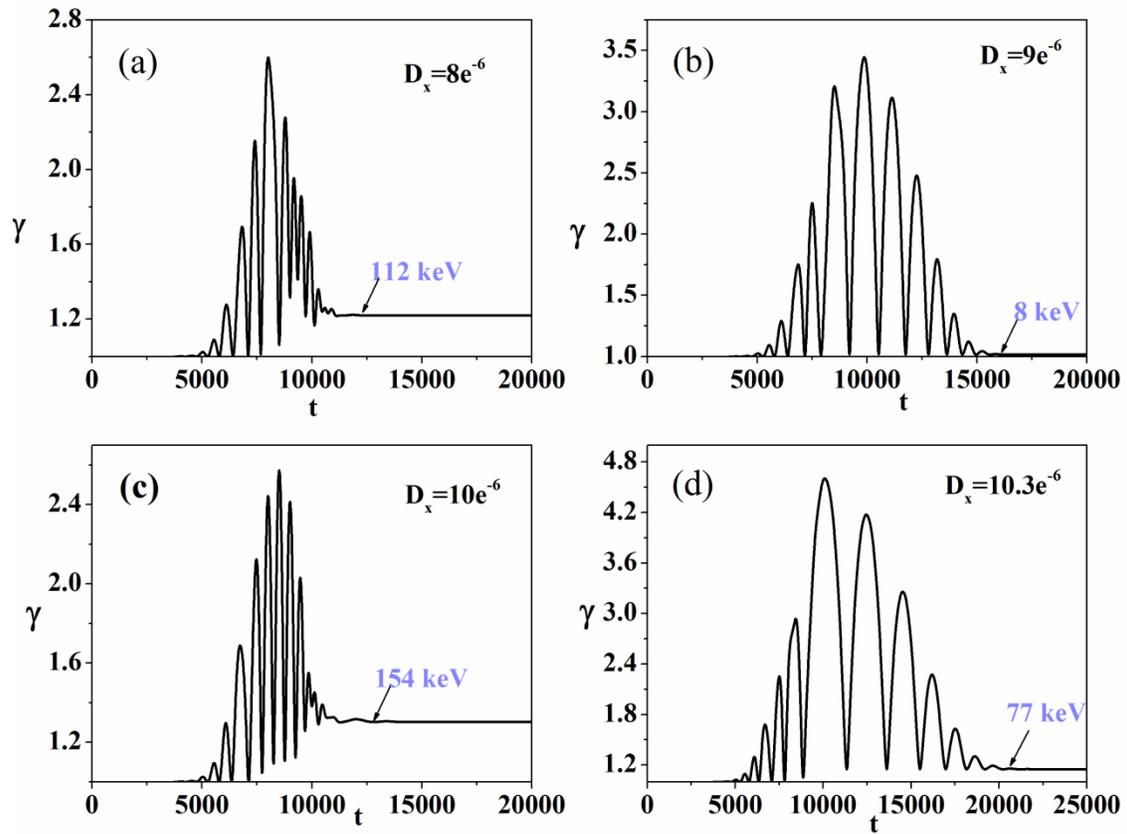

**Fig. 3**

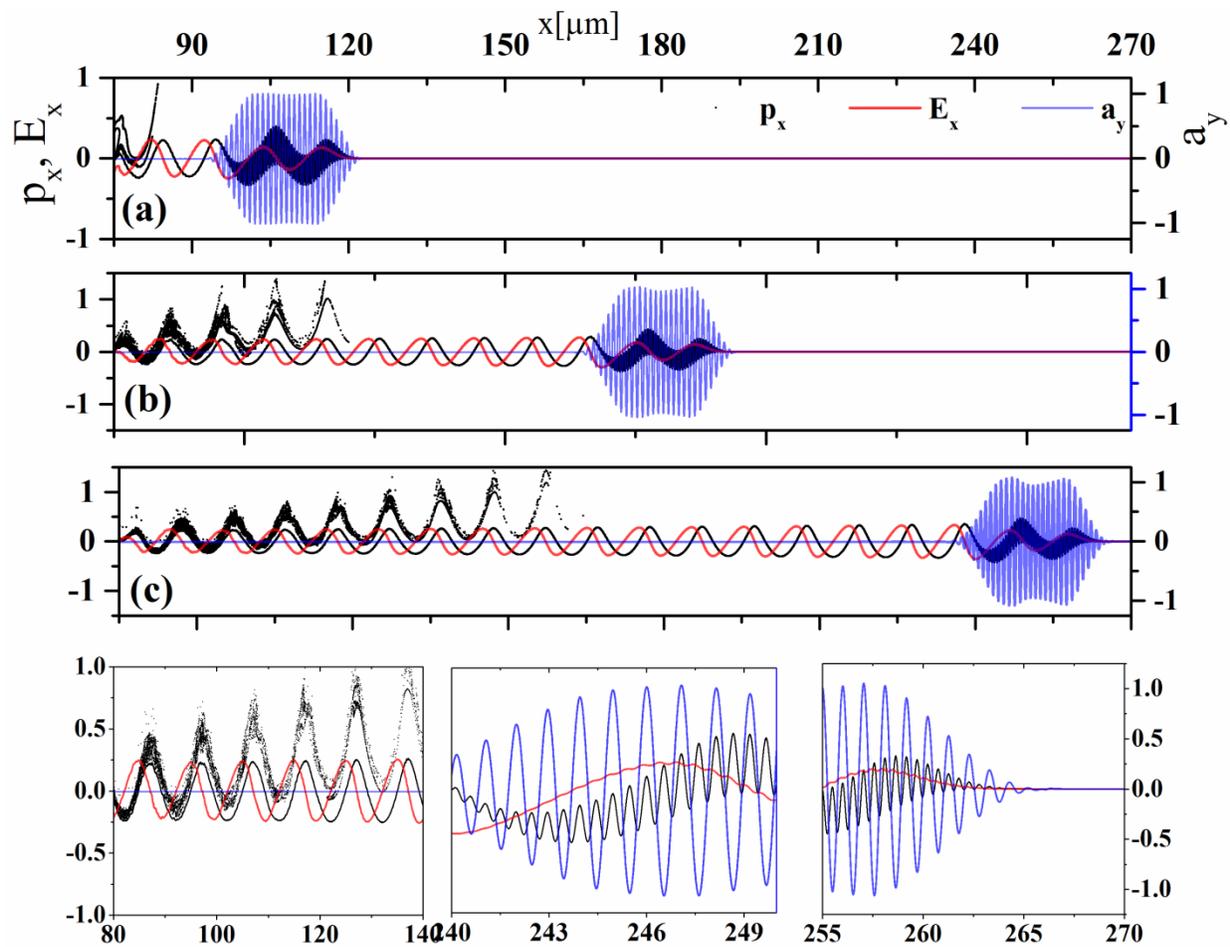

**Fig.4**

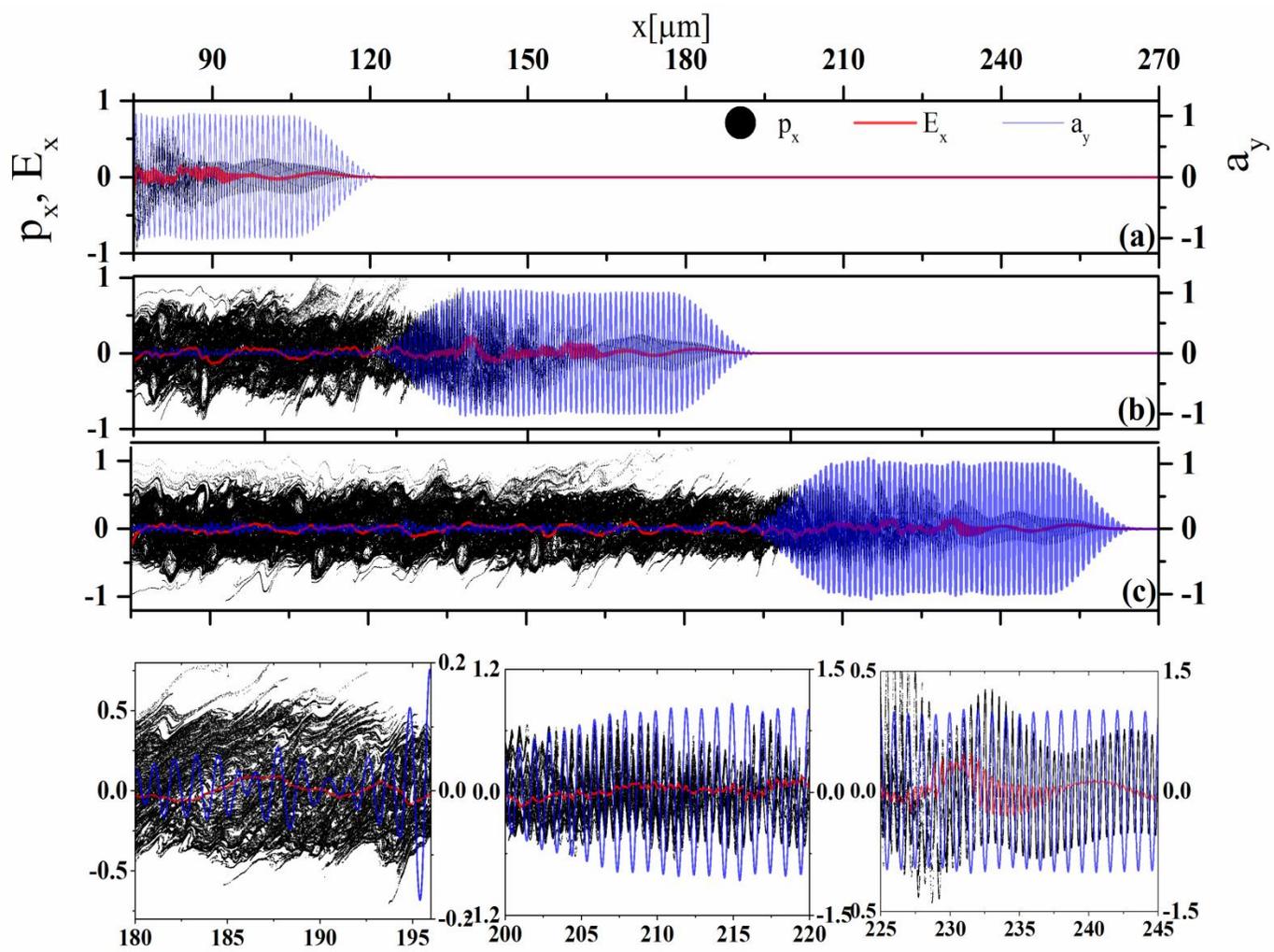

**Fig.5**

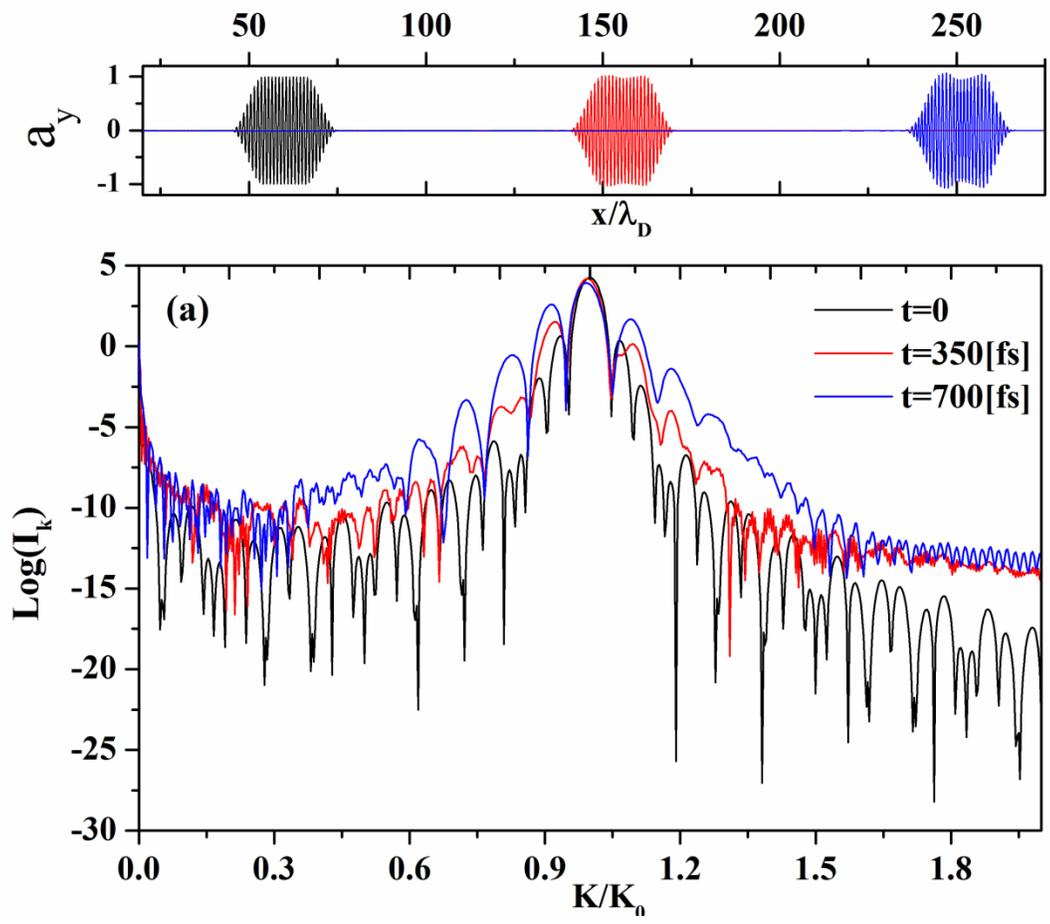

**Fig. 6 (a)**

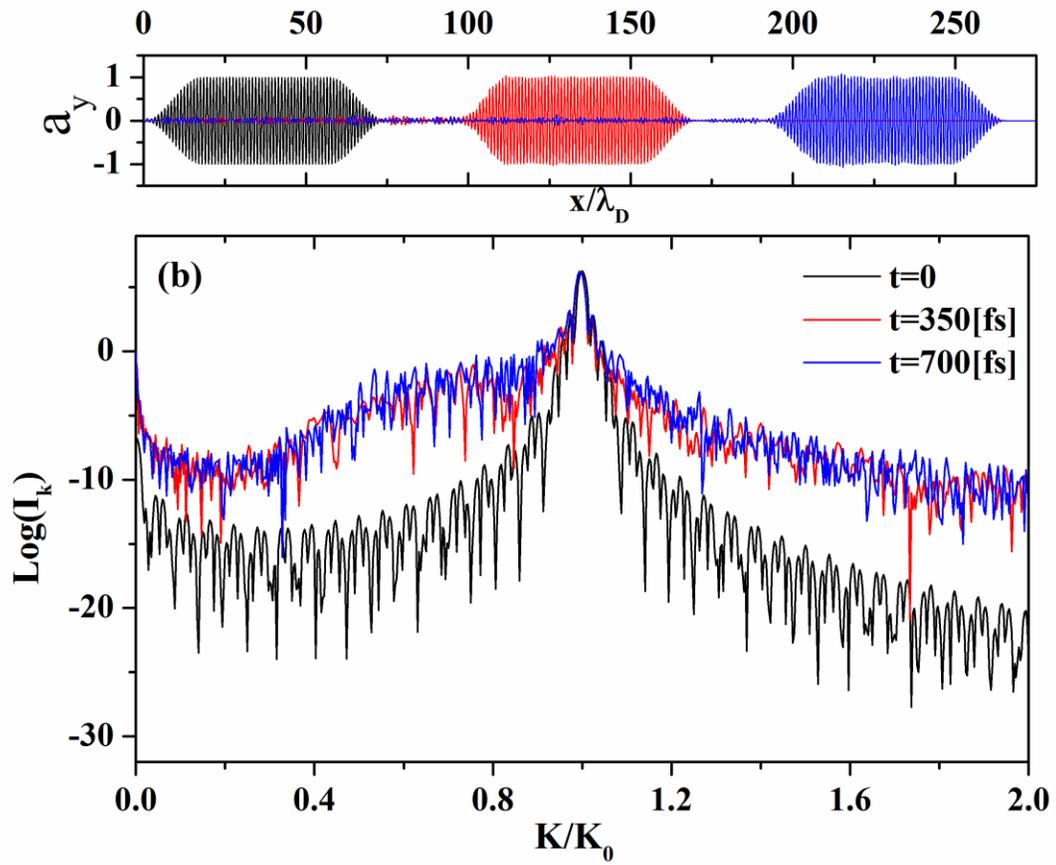

**Fig. 6 (b)**

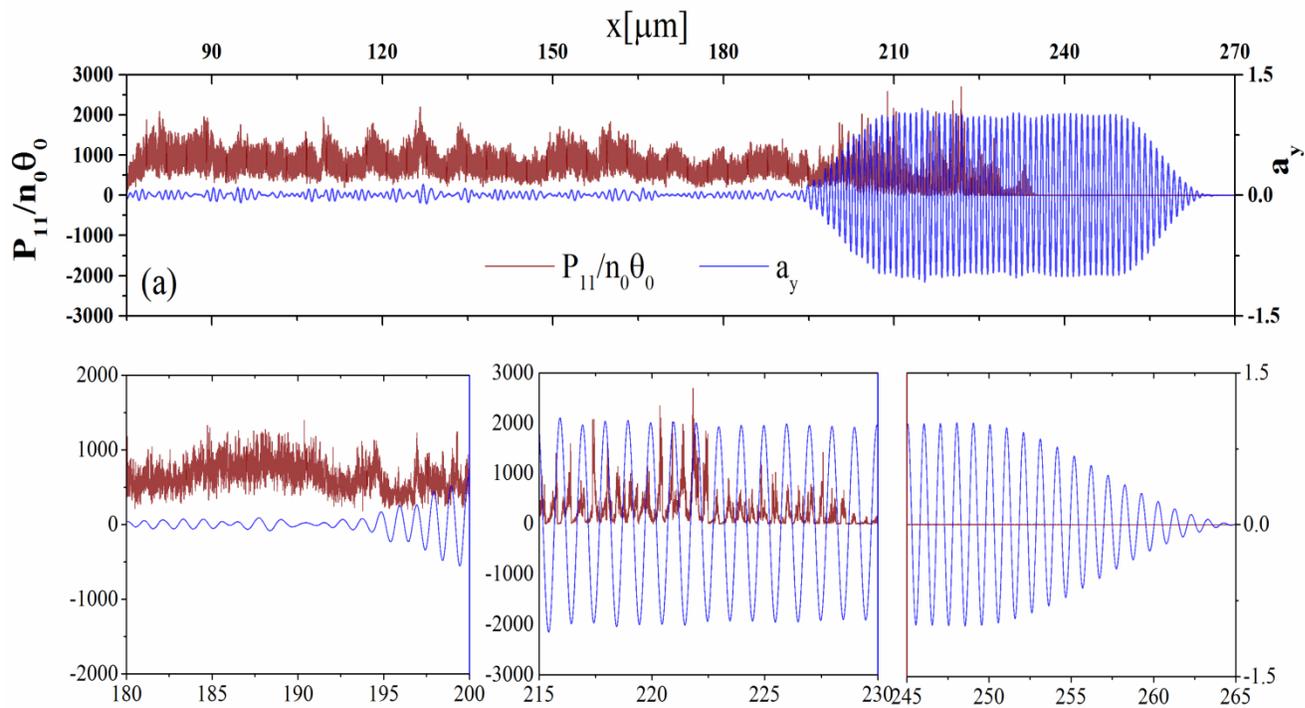

**Fig.7 (a)**

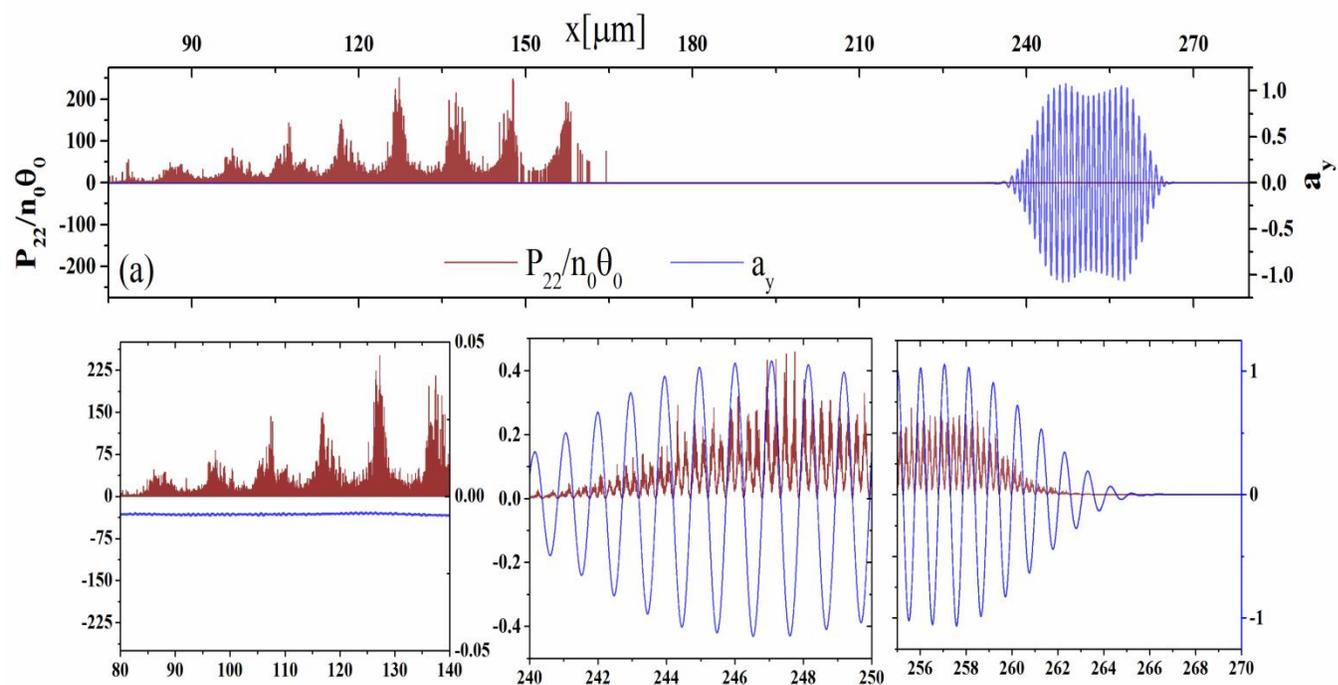

**Fig. 7 (b)**

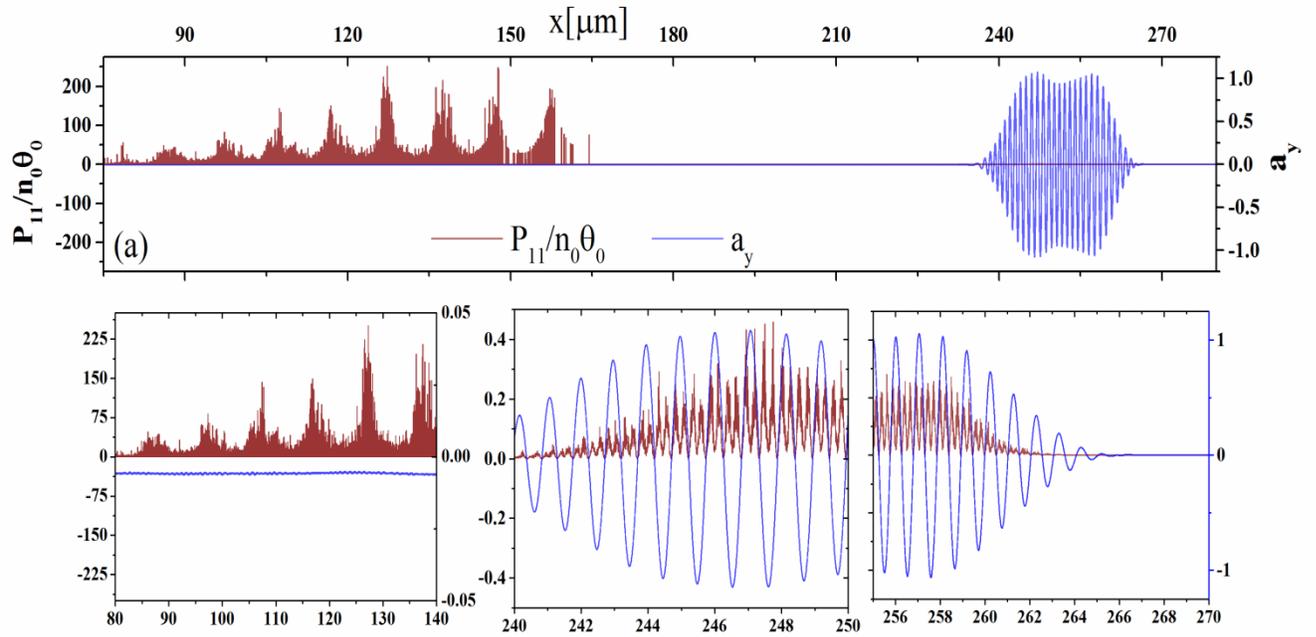

**Fig. 8 (a)**

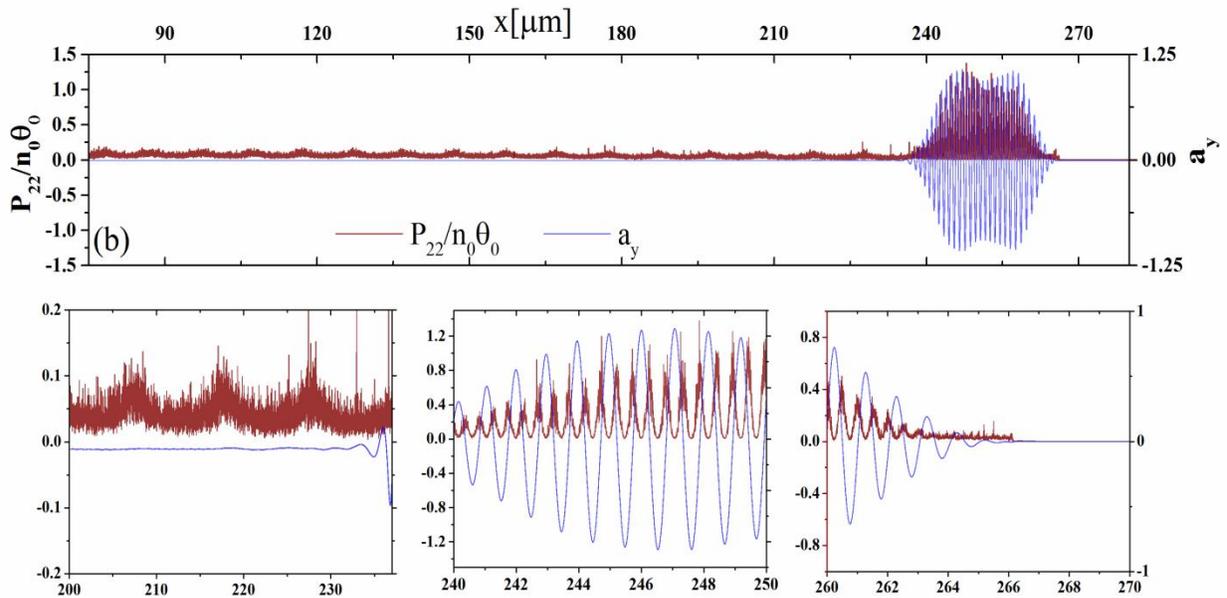

**Fig. 8 (b)**